\documentclass[fleqn,twoside]{article}

\usepackage{graphicx}
\usepackage[figuresright]{rotating}
\usepackage{amsmath}
\usepackage{amssymb}
\usepackage{espcrc2}
\usepackage{mcite}

\def\sign#1{\hbox{\rm \,sign}(#1)}

\title{Improving Inversions of the Overlap Operator}

\author{S. Krieg\address[PW]{Department of Physics, Universit\"at
    Wuppertal, Gaussstrasse 19, Wuppertal, Germany.}\thanks{Based on
    poster presented by S. Krieg at lattice 2004.},
        N. Cundy\addressmark[PW]
\thanks{NC is supported by EU 
        grant MC-EIF-CT-2003-501467}, %
        J.~van den Eshof\address[MW]{Department of Mathematics , University of D\"usseldorf, Germany.},
        A. Frommer\address[MW]{Department of Mathematics, Universit\"at
    Wuppertal.},
	Th. Lippert\address[J]{John von Neumann Institute for Computing,
	  J\"ulich Research
	  Centre, 52425 J\"ulich, Germany.},
	and K. Sch\"afer\addressmark[MW]. 
}
       
\begin{document}

\begin{abstract}
{We present relaxation and preconditioning techniques which accelerate
  the inversion of the overlap operator by a factor of four on small
  lattices, with larger gains as the lattice size increases. These
  improvements can be used in both propagator calculations and dynamical simulations.}
\end{abstract}

\maketitle
\section{INTRODUCTION}
The massive overlap operator
can be written as
\[
D_u = \rho I + \gamma_5 \mathrm{sign}(Q),\nonumber
\]
where $\rho \ge 1$ is a mass parameter corresponding to a bare fermion mass
$m_b = m(\rho - 1)$, and $Q$ is the hermitian Wilson operator with a
negative mass parameter $-m$. Note that $\gamma_5\sign Q$ is unitary. We can also define a
hermitian overlap operator $D_h = \gamma_5 D_u$. Both
operators are normal, i.e.\ they commute with their adjoints. We will approximate the
matrix sign function using the Zolotarev partial fraction expansion (ZPFE),
\[
\mathrm{sign}(Q) = \sum_{i = 1}^{N_Z}\frac{\omega_i Q}{Q^2 + \tau_i},\nonumber
\]
where the coefficients $\omega_i$ and $\tau_i$ are known \cite{paper1}, and
the accuracy of the approximation depends on the Zolotarev order $N_Z$.
Multiplication of the ZPFE with a vector is easily implemented using a multishift CG-inversion \cite{mmass}.

In these proceedings, we will sketch how to optimise the inversion of $D_u$ (the
propagator calculation), and $D_h^2$ (the full inversion). Both these
inversions can be reduced to solving for the vector $x$ in any of the
three following equations:
\begin{eqnarray}
D_u x = & b,\label{eq:1}\\
D_h x = & \gamma_5 b,\label{eq:2}\\
D_h D_h x = &b.\label{eq:3}
\end{eqnarray}

\section{SUMR}

\begin{figure}
\begin{tabular}{c}
\includegraphics[width =\columnwidth,height = 5cm]{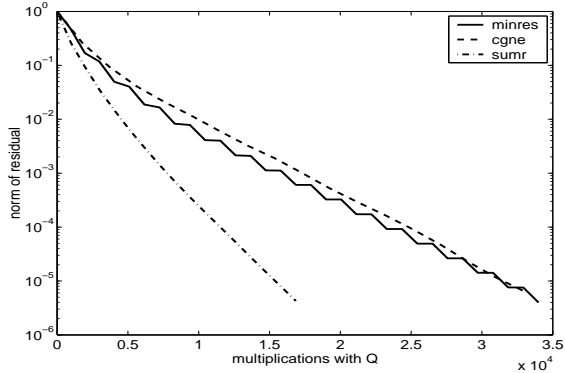}
\end{tabular}
\caption{The residual of a SUMR inversion compared to CG and MINRES on an
  dynamical $8^4$ configuration, plotted against calls to the Wilson operator.}\label{fig:fig1}
\end{figure}

The optimal method to solve equation (\ref{eq:2}) is well known to be
the minimal residual (MINRES). Similarly, for equation
(\ref{eq:3}) the conjugate gradient (CG) method is optimal. Less well
known is the shifted unitary minimal residual (SUMR) method
\cite{JagelsReichel,paper2}, which can be
used to invert equation (\ref{eq:1}). There are known formulae for the accuracy of
the inversion for each of these methods after $k$ iterations,
which can be used to calculate a worst case bound for the number of iterations $k(\epsilon,\rho)$
needed to achieve an accuracy $\epsilon$ \cite{paper2}. $x^\ast$ is
the true solution, $x^k$ the approximate inverse, and $r^k = D_u x^k -
b$ is the residual.
\begin{itemize}
\item[(i)] \ For SUMR we have
\[ \parallel x^k - x^\ast \parallel_2 \, \leq
\frac{1}{\rho - 1} \parallel r^k_u \parallel_2 \, \leq \frac{2}{\rho - 1}
\left( \frac{1}{\rho}\right)^k \parallel r^0 \parallel_2,
\]
\[
\Rightarrow k (\varepsilon, \rho) \, \leq \frac{-\ln ( \varepsilon )}{\ln (\rho)} + \frac{-
\ln (2/(\rho - 1))}{\ln (\rho)} .
\]
\item[(ii)] \ For MINRES we use $\|r_h^0\|_2 = \|r_u^0\|_2$, since $r_h^0 = \gamma_5 r_u^0$,
\[
\parallel x^k - x^\ast \parallel_2 \, \leq
\frac{1}{\rho - 1} \parallel r^k_h \parallel_2 \, \leq \frac{2}{\rho - 1}
\left( \frac{1}{\rho}\right)^{\lfloor \frac{k}{2}  \rfloor}
\parallel r^0_h \parallel_2,
\]
\[
\Rightarrow k (\varepsilon, \rho) \, \leq 2 \left( \frac{-\ln (\varepsilon)}{\ln
(\rho)} + \frac{-\ln (2/(\rho - 1))}{\ln (\rho)} \right) .
\]

\item[(iii)] \ For CG (eqn.~(\ref{eq:3}), with two calls to $D_h$ per iteration) we have
\[
\parallel x^k - x^\ast \parallel_2
\leq 2 \left( \frac{1}{\rho} \right)^k \parallel x^0-x^\ast\parallel_2
\leq
\frac{2}{\rho^k(\rho -
1)} \parallel r^0 \parallel_2 ,
\]
\[
\Rightarrow k (\varepsilon, \rho) \, \leq \frac{-\ln  (\varepsilon)}{\ln (\rho)} + \frac{ -
\ln (2/(\rho - 1))}{\ln (\rho)} .
\]
\end{itemize}
{Based on these worst case estimates, SUMR is a factor 2 faster than
  CG or MINRES for a propagator calculation. We see this factor of 2
improvement numerically (see figure \ref{fig:fig1}).}

\section{RELAXATION}


While solving $D_u x = b$, there is no need to calculate the sign function
to the full accuracy ($\equiv\epsilon$) during the entire inversion. For each step of the
inversion we have to calculate an approximation $s^{j}$ to the product
of the overlap operator with a vector
$y^j$. Here $\eta^j$ is the relative accuracy of the sign function:
$$
\|D_u y^j - s^j\| \le \eta_j \cdot \|D_u\| \cdot \|y^j\|.
$$
The precision $\eta$ can be fixed at a value less than $\epsilon$, but it
is more efficient to increase $\eta$ as the inversion progresses. The key
is to ensure that the residual gap
\[
\begin{array}{clcl}
\|\underbrace{b-Ax^k}\| &\le& \|\underbrace{r^k - (b-Ax^k)}\| &\nonumber\\
\hbox{true residual} & & \hbox{residual gap} & \\
&+&\|\underbrace{r^k}\|&\\
&&\hbox{computed residual}&
\end{array}
\]
{{at the end of the calculation is of order $\epsilon$. The optimal
relaxation strategies are \cite{paper2}:}}\\
\vspace*{0.5\baselineskip}\\\vspace*{0.5\baselineskip}
{
\begin{tabular}{ll}
\hline
 method & tolerance $\eta_j$  \\
\hline
 CG & $\eta_j = \epsilon \sqrt{\sum_{i=0}^j \|r^i\|^{-2}}$  \\
 MINRES & $\eta_j = \epsilon/\|r^j\|$  \\
SUMR & $\eta_j = \epsilon/\|r^j\|$ \\
\hline
\end{tabular}
}\\
{The accuracy of the sign function can be relaxed by reducing the
accuracy of the multishift solver used to solve the ZPFE, and by reducing $N_Z$. This
relaxation can be applied to CG (we call the relaxed CG inversion
``relCG'') and SUMR (``relSUMR''). Relaxation gains a factor of 1.5-1.8 in
computer time (see section \ref{sec:JVE}).}

\section{JVE PRECONDITIONING}\label{sec:JVE}

\begin{figure}
\begin{tabular}{c}
\includegraphics[width =\columnwidth,height = 5cm]{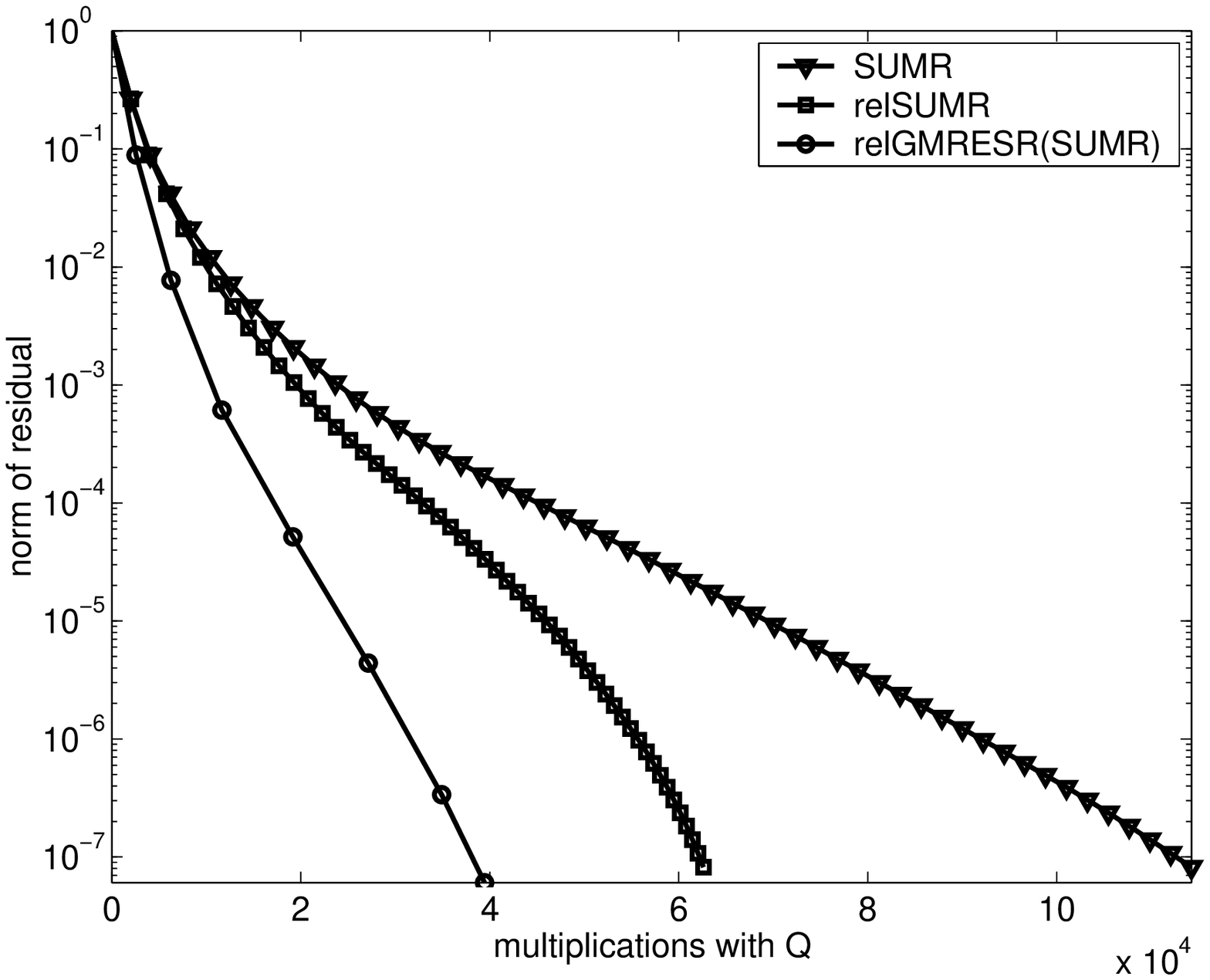}\\
\includegraphics[width =\columnwidth,height = 5cm]{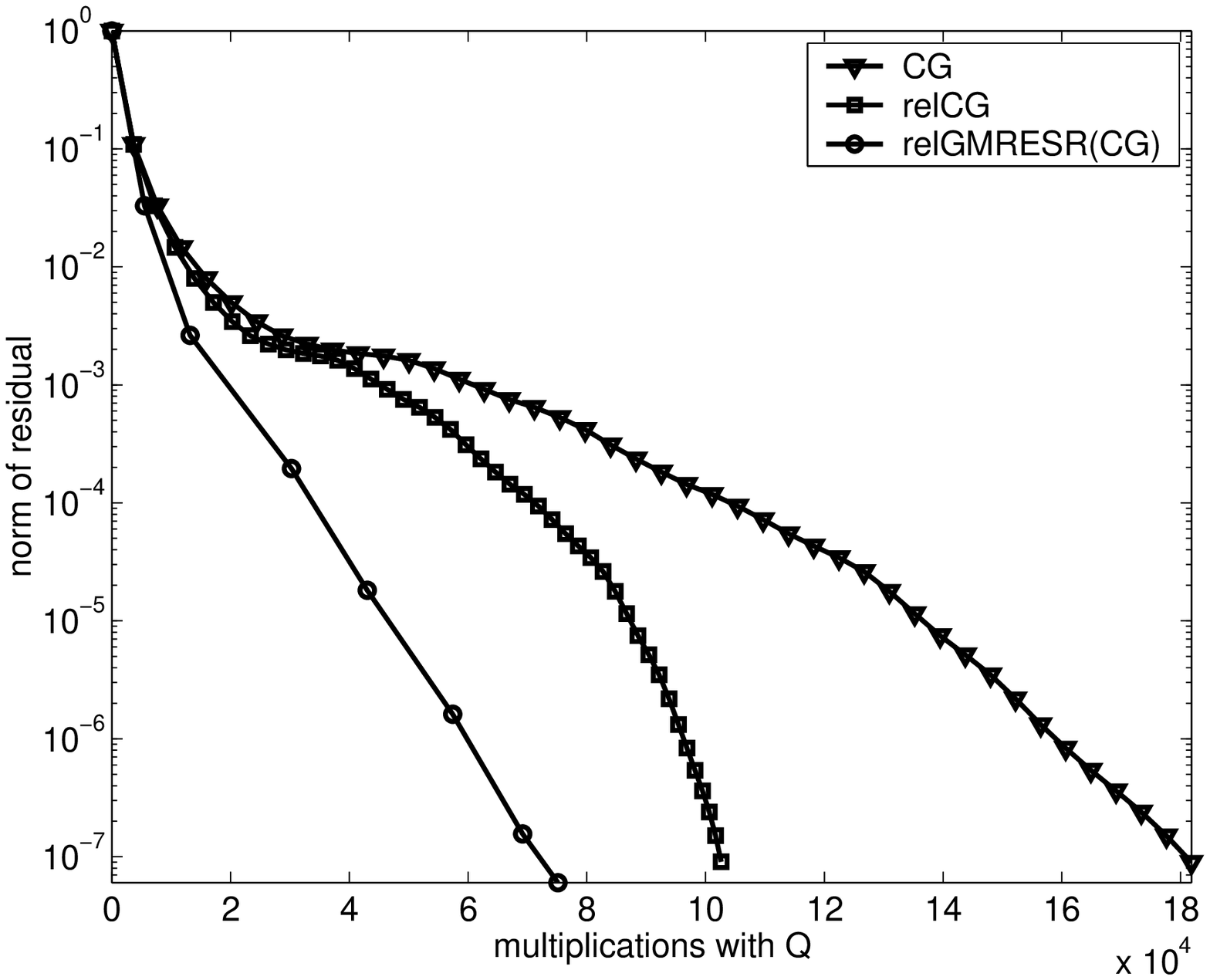}
\end{tabular}
\caption{The norm of the residual compared to calls to the Wilson
  operator when solving (\ref{eq:1}) using SUMR, relSUMR and
  GMRESR(SUMR) (top), or when solving (eqn.~(\ref{eq:3})) using CG,
  relCG and relGMRESR(CG). The calculations were done on an $8^4$ configuation with $\rho = 1.86$ }\label{fig:fig2}
\end{figure}

We can achieve further gains by using inversion of a low-accuracy
overlap operator as a preconditioner; we use relSUMR (as preconditioner for the
propagator inversion) or relCG (for the full inversion) with the relative
accuracy $\epsilon = 0.01$, and $N_Z = 5$ (these numbers can be optimized).
This preconditioner can be used in an inversion algorithm which can
support a variable preconditioner: we used the GMRESR algorithm
\cite{paper3, original_GMRESR_paper}. The GMRESR inversion can be relaxed,
using the methods of the previous section, so we have chosen to call this
inversion algorithm relGMRESR(SUMR) for the propagator inversion or
relGMRESR(CG) for the full inversion. This preconditioning achieves a gain
of at least a factor of 4 in computer time (see figure \ref{fig:fig2}
and tables \ref{tab:tab1} and \ref{tab:tab2}).\\

\begin{table}
\begin{tabular}{l|l l l}
\hline
Method&$\rho = 1.86$&$\rho = 1.22$\\
\hline
CG            &1258&2019
\\
relCG         &729(1.72)&1229(1.64)
\\
relGMRESR(CG) &312(4.04)&576(3.51)
\\
\hline
SUMR           &1313&2489
\\
relSUMR        &782(1.68)&1619(1.54)
\\
relGMRESR(SUMR)&372(3.52)&633(3.93)
\\
\hline
\end{tabular}
\caption{Timings for one inversion of (\ref{eq:3}), or two inversions
  of (\ref{eq:1}) averaged over three dynamical $\beta = 5.6$ $8^4$
  configurations. The number in brackets is the gain from the
  unrelaxed preconditioned case.}\label{tab:tab1}
\end{table}

\begin{table}
\begin{tabular}{l | l l}
\hline
Method&$\rho = 1.22$&$\rho = 1.06$
\\
\hline
CG             &9022&31430
\\
relCG          &5981(1.51)&18813(1.67)
\\
relGMRESR(CG)  &2329(3.87)&6642(4.73)
\\
\hline
SUMR           &8312&31550
\\
relSUMR        &6038(1.38)&18840(1.87)
\\
relGMRESR(SUMR)&2252(3.69)&5974(5.82)
\\
\hline
\end{tabular}
\caption{Same as table \ref{tab:tab1}, but on a quenched $\beta = 6.0$
  $16^4$ configuration.}\label{tab:tab2}
\end{table}

\section{CONCLUSIONS}

We present algorithms which accelarate the invesion of the overlap operator by a factor of about
four for the full inversion (eqn. (\ref{eq:3})). Using
relGMRESR(SUMR) can give a gain of a factor of at least 10 over MINRES for the
propagator calculation. We expect that the gain will be increased on larger
systems and at lower masses.


\bibliographystyle{elsart-num} 
\bibliography{proceedings}

\end{document}